\documentclass[oldversion]{aa}
\usepackage{graphics,natbib}
\bibpunct{(}{)}{;}{a}{}{,}

\begin{document}

\title{Observations and modelling of a clumpy galaxy at $z=1.6$
\thanks{Based on observations obtained at the Very Large Telescope (VLT) of the European Southern Observatory, Paranal, Chile (ESO program ID 278.A-5009)}}

\subtitle{Spectroscopic clues to the origin and evolution of chain galaxies}

\author{F. Bournaud \inst{1}, E. Daddi \inst{1}, B. G. Elmegreen \inst{2}, D. M. Elmegreen \inst{3}, N. Nesvadba \inst{4},\\
E.~Vanzella \inst{6},  P.~Di~Matteo\inst{4,5}, L. Le~Tiran \inst{4}, M. Lehnert \inst{4}, D. Elbaz \inst{1}}

\offprints{F. Bournaud \email{frederic.bournaud@cea.fr}}

\institute{Laboratoire AIM, CEA-Saclay DSM/IRFU/SAp -- CNRS -- Universit\'e Paris Diderot, 91191 Gif-sur-Yvette, France
\and
IBM Research Division, T.J. Watson Research Center, P.O. Box 218, Yorktown Heights, NY 10598, USA
\and
Vassar College, Dept. of Physics \& Astronomy, Box 745, Poughkeepsie, NY 12604, USA
\and
Observatoire de Paris, GEPI, F-92195 Meudon, France
\and
Observatoire de Paris, LERMA, F-75014 Paris, France
\and
INAF -- Osservatorio Astronomico di Trieste, Via G.B. Tiepolo 11, 40131 Trieste, Italy
}

\date{Received; accepted}

\abstract{
We investigate the properties of a clump-cluster galaxy at redshift 1.57. The morphology of this galaxy is dominated by eight star-forming clumps in optical observations, and has photometric properties typical of most clump-cluster and chain galaxies. Its complex asymmetrical morphology has led to the suggestion that this system is a group merger of several initially separate proto-galaxies. We performed H$\alpha$ integral field spectroscopy of this system using SINFONI on VLT UT4. These observations reveal a large-scale velocity gradient throughout the system, but with large local kinematic disturbances. Using a numerical model of gas-rich disk fragmentation, we find that clump interactions and migration can account for the observed disturbed rotation. On the other hand, the global rotation would not be expected for a multiply merging system. We further find that this system follows the stellar mass vs. metallicity, star formation rate and size relations expected for a disk at this redshift, and exhibits a disk-like radial metallicity gradient, so that the scenario of internal disk fragmentation is the most likely one. A red and metallic central concentration appears to be a bulge in this proto-spiral clumpy galaxy. A chain galaxy at redshift 2.07 in the same field also shows disk-like rotation. Such systems are likely progenitors of the present-day bright spiral galaxies, forming their exponential disks through clump migration and disruption and fueling their bulges. Our present results show that disturbed morphologies and kinematics are not necessarily signs of galaxy mergers and interactions, and can instead result from the internal evolution of primordial disks.
}
\authorrunning{Bournaud et al.}
\titlerunning{Kinematics of a clumpy galaxy at $z=1.6$ and the nature of chain galaxies}
\maketitle


\section{Introduction}
Giant molecular clouds and star formation complexes in present-day spiral galaxies have masses that rarely exceed $10^{-3}$ of the disk mass, and the total mass in these low-mass star forming regions does not dominate the disk mass. At high redshift ($z \simeq 1$ and above) the clumpiness and asymmetry of galaxies increases \citep{conselice03, conselice04,daddi04}, which affects both the light fraction that is in the clumps and the individual masses of these clumps. Clumps at high redshift can be kpc-wide and as massive as $10^9$~M$_{\sun}$ \citep[][and references therein]{E07IAU}. In particular, the so-called ''chain galaxies'' have striking morphologies dominated by alignments of massive clumps \citep{cowie96,vdbergh96,moustakas04}. \citet{dalcanton96} first proposed that these chain galaxies could be edge-on low surface brightness galaxy (LSB) progenitors at high redshift, but their clumps are one order of magnitude brighter than the star-forming regions of LSBs \citep{smith01} and the clumps are not only bright star-forming complexes but are massive structures also visible in the near infrared (\citealt[][hereafter EE05]{EE05}, see also NICMOS observations by \citealt{dickinson2000}). ''Clump cluster'' galaxies have the appearance of highly clumpy disks. Some are bulgeless, others have small red central concentrations that resemble primordial bulges. Because the photometrical properties of their clumps are similar to that of chain galaxies, \citet{EEH04} suggested that chains are the edge-on counterpart of such clumpy systems, which solves the problem of the face-on counterpart of chains being missing \citep{oneil00}. The fraction of chains and clump clusters increases with redshift, and because these galaxies do not have an exponential profile already established, \citet{EEVFF} proposed that they could be the progenitor of $z \sim 1$ spirals that further evolved into present-day bright spirals. They would thus be a crucial step in the formation of massive disk galaxies.

The origin of the massive clumps in chains and clump cluster galaxies, however, remains uncertain and largely debated. \citet{noguchi} and \citet{immeli1,immeli2} proposed that these could be dense fragments resulting from Jeans instabilities in gas-rich and bulgeless disks. \citet*[][hereafter BEE07]{BEE07} further explored this hypothesis and confirmed that these would evolve into spiral disks with a central bulge and an exponential disk (generally truncated in the form of type II as defined in \citet{pohlen07}). This hypothesis of internal instability however lacks clear support from observations so far. Spectroscopic observations of high-redshift disks unveil high turbulent speeds \citep{genzel06,FS06} so that the Jeans mass should be high and massive clumps could be formed, but these extensive spectroscopic surveys lack associated deep imaging to directly adress the origin of clumpy galaxies. In contrast, \citet{taniguchi} favored a multiple-merger origin involving several initially independent proto-galaxies. In that vein, clump-clusters could be the high redshift equivalent of local compact groups, and some indeed have Stefan's Quintet-like morphology (at the resolution of high-reshift observations). The merging subunits could be aligned along filamentary structures in the case of chains \citep{taniguchi}. Even binary major mergers can look clumpy with several star forming knots \citep{goldader}, and even an exponential profile would not exclude a merger origin \citep{moran}. Complex morphologies with severe disturbances and asymmetries, like for instance bent chains or shrimp-like systems \citep{EE06,EEFM07} further suggest that and interaction/merger origin is plausible at least for these categories. 

In this paper, we present kinematical IFU observations of the prototype clump-cluster galaxy UDF~6462 in the Hubble Ultra Deep Field (UDF). The morphology and photometry of this system at $z=1.57$ have been studied in EE05 and \citet{EERC07}. A color image and an HST/ACS V-band image from the Hubble Ultra Deep Field \citep[UDF,][]{udfpaper} are presented in Fig.~1 and 2, the main photometrical properties are recalled in Sect.~\ref{21}. This system is one that a priori poses a problem for the hypothesis of clumps being internally formed by instabilities in a massive disk: it is globally asymmetrical and has a Southern extension that resembles a spiral arm. Furthermore, two nearby systems (in projection) could be involved in an on-going interaction. UDF~6462 is classified as an on-going merger by \citet{conselice07}. 

Integral field spectroscopy data obtained with SINFONI on UT4 at ESO/VLT reveal complex internal kinematics, which has local disturbances but is overall dominated by a velocity gradient resembling that expected for an internally rotating system. Using the numerical simulations presented in BEE07, we propose a model of an unstable lopsided disk which forms internal clumps and, under some projections, a morphology resembling that of UDF~6462. We show that the observed velocity gradient and kinematical disturbances correspond precisely to those expected in such a scenario. Thus, this object that could a priori be regarded as being an on-going merger prototype, actually has properties that are fully compatible with the internally unstable disk hypothesis. Furthermore, the kinematics and some other properties are more difficult to explain in a multiple-merger scenario. Kinematical results of another nearby clumpy galaxy with a bent-chain morphology (UDF~6911) at $z \simeq 2$ are also presented, and also show a large-scale velocity gradient favouring an internal disk fragmentation rather than a merger origin.

SINFONI observations of UDF~6462 are presented in Section~2. In Section~3, we present the numerical model that helps to interpret the morphological and kinematical properties of this system. In Section~4, we discuss the origin and evolution of this clump cluster galaxy, and our conclusions are summarized in Section~5. Spatially resolved SEDs in this system are presented in Appendix~A ; numerical models of metals distribution in disks and mergers are discussed in Appendix~B.
In this paper we use a \citet{chabrier} IMF and standard cosmological parameters with $H_0=75$~km~s$^{-1}$~Mpc$^{-1}$. Throughout the paper, we will call {\it clumps} the large and dense structures typical of high-redshift clump-cluster galaxies, that have masses of $10^{8-9}$~M$_{\sun}$ and typical sizes of the order of a kpc, much more massive than kpc-sized star-forming ''clumps'' of low-redshift spiral galaxies.


\section{Observations}
\subsection{The UDF~6462 clump cluster}\label{21}

UDF~6462 (Fig.~1) has been imaged in $B$, $V$, $i$ and $z$ bands using HST/ACS and $J$ and $H$ bands with NICMOS. The main properties are summarised in Table~\ref{tab1}. A spectroscopic redshift of $z=1.570$ has been measured during the FORS2 campaign on the GOOD-S field \citep{vanzella06,vanzella08} based on detection of the [OII] line. $BViz$ data from \citet{giavalisco} are presented and analyzed by EE05 and a photometrical analysis including NIR bands is in Appendix~A. We here show V- and H-band images in Fig.~\ref{f2} to recall the morphology of the system. EE05 estimate a stellar mass of $3.3 \times 10^{10}$~M$_{\sun}$ and have identified 8 star-forming clumps in this system, with 36\% of the total $i_{775}$ luminosity in these clumps and 64\% in a diffuse and redder inter-clump medium. The typical age of the clumps derived from optical photometry is around 310~Myr, while the inter-clump medium has a typical stellar age of nearly 3~Gyr. These properties, as well as the color-magnitude properties of the clumps, are globally typical of the sample of ten clump-cluster galaxies studied by EE05. The clump-cluster also has a central reddish blob, and the NIR emission is much more concentrated than the optical emission; this resembles a primordial bulge. In the following we will call this structure {\it "bulge"} and subsequently in the paper we will make the case for this; for the moment we only imply this word to be indicative of a location within the system and not presuming of its real nature.\\
The $B-i$ ACS image in Fig.~\ref{f3} shows a color gradient from this red central proto-bulge to the bluer clumps. This is not simply reddening caused by higher extinction, considering that the region with red $B-i$ colors corresponds to the area of highest H-band surface brightness (see also SED fits in Appendix~A). Photometric measurements compared with stellar population models suggest a bulge mass fraction of about 15--20\%. This is actually an upper limit because of the NICMOS H-band resolution being larger than the red blob diameter. Spectral energy distributions of this central bulge-like blob and the other clumps are shown in Appendix~A and also discussed in Section~4.

\begin{figure}
\centering
\resizebox{8cm}{!}{\includegraphics{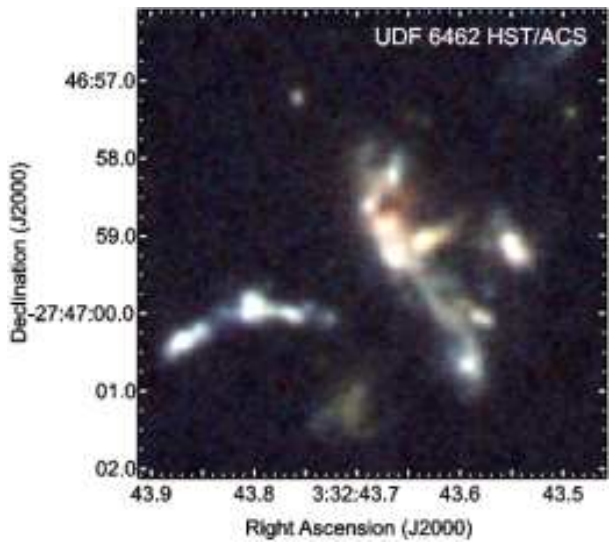}}
\caption{Color image of the UDF~6462 clump-cluster from optical multi-band ACS imaging. The bent-chain UDF~6911 is visible to the left.}\label{f1}
\end{figure}

\begin{table}
\centering
\begin{tabular}{lc}
\hline
\hline
\multicolumn{2}{c}{UDF~6462}\\
\hline
$\alpha$ (J2000) &  3$h$ 32$m$ 43.65$s$  \\
$\delta$ (J2000)  &  -27$^o$ 46$'$ 59.2$''$ \\
redshift &  1.571 \\
$B$ mag & 24.16 \\
$i$ mag & 23.60 \\
M$_*$ & $3.3 \times 10^{10}$~M$_{\sun}$ \\
SFR$_\mathrm{UV}$ & 50~M$_{\sun}$~yr$^{-1}$  \\
SFR$_\mathrm{IR}$ & 85~M$_{\sun}$~yr$^{-1}$ \\
12+log(O/H) & $8.53 \pm 0.02$ \\
V$_{\mathrm{circ}}$ & $\sim 100$~km~s$^{-1}$\\
\hline
\end{tabular}
\caption{Main properties of UDF~6462. Details and references are in Section~2.1.}\label{tab1}
\end{table}

\begin{figure*}
\centering
\resizebox{18cm}{!}{\includegraphics{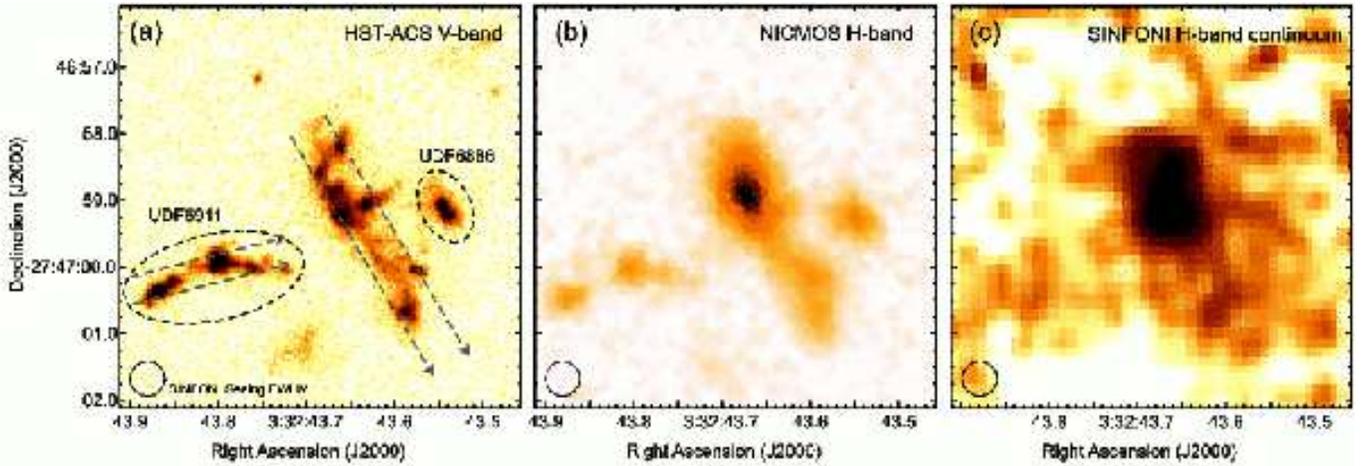}}
\caption{The UDF~6462 clump-cluster: {\bf (a)} HST-ACS V-band image plotted in log-scale. UDF~6911 and 6886 are are very different redshifts (see text) and not interacting with UDF~6462. Dashed arrows indicate the pseudo-slits used to derive position-velocity diagrams.  {\bf (b)} NICMOS H-band image plotted in log-scale, showing a more concentrated emission around the central bulge-like blob. {\bf (c)} H-band continuum from our SINFONI observations plotted in log-scale: UDF~6462 is detected, including its faint southern extensions; UDF~6911 is also detected.}\label{f2}
\end{figure*}

UDF~6462 has two neighboring systems, which we here denote UDF~6911 and UDF~6886 (see labelling in Fig.~\ref{f2}). UDF~6911 has a chain-like morphology, while UDF~6886 is more compact. The colors and dimensions are comparable to that of UDF~6462, possibly suggesting a similar redshift. We actually find from our SINFONI observations (see below) that these are at different redshifts.

The star formation rate (SFR) derived from the Spitzer MIPS 24~$\mu$m flux (Chary et~al. in preparation) is 85~M$_{\sun}$~yr$^{-1}$ based on the \citet{charyelbaz} models. This value could be an over-estimate because the 24~$\mu$m flux is likely contaminated by nearby systems. The UV-based estimate corrected for extinction is 50~M$_{\sun}$~yr$^{-1}$, as in \citet{daddi07}, and is not contaminated by nearby systems. The [OII] flux, corrected for extinction, corresponds to an SFR of 35--55~M$_{\sun}$~yr$^{-1}$ (using relations from Kennicutt~1998a). As the  24~$\mu$m flux may be contaminated by nearby systems (in particular UDF~6911 and 6886), these various estimates are consistent and suggest an SFR of about 50~M$_{\sun}$~yr$^{-1}$ with an uncertainty around 20~M$_{\sun}$~yr$^{-1}$.


\begin{figure}
\centering
\resizebox{8cm}{!}{\includegraphics{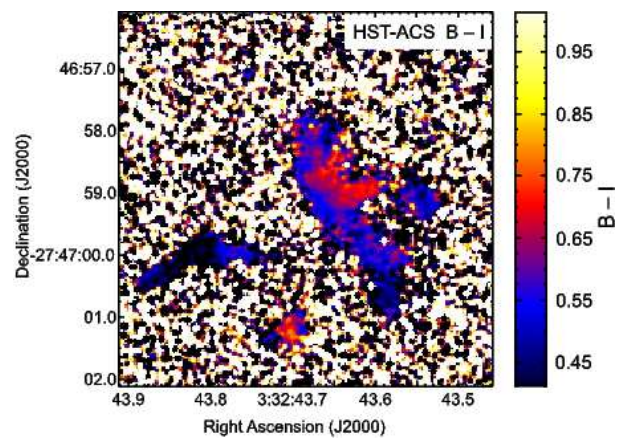}}
\caption{$B-i$ color map of UDF~6462 from HST/ACS data, showing the color gradient from the central proto-bulge to the outer clumpy regions.}\label{f3}
\end{figure}

\subsection{SINFONI observations and data reduction}

The observations were carried out using the SINFONI near-infrared integral field spectrometer mounted on the VLT UT4 telescope \citep{eisenhauer, bonnet, gillessen}. An integration time of 5 hours was spent on the target in December 2006 with average seeing of 0.5 to 0.6$\arcsec$ in the H-band. No adaptative optics was used because no suitable Natural Guide Star was found close enough to the target. We used the high-resolution H-band grism, which offers a FWHM spectral resolution of $R \simeq 3000$ in the H-band, equivalent to a FWHM of $\sim 90$~km~s$^{-1}$ for the H$\alpha$ line at $z=1.57$. A first standard reduction of the data has been performed using the SINFONI pipeline \citep{sinfopipe}. 

We then performed a more careful reduction relying on a set of standard IRAF tools to reduce longslit
spectra \citep{tody93}, which we modified and extended by a set of IDL
routines matched to the special requirements of SINFONI. The dark-frame
subtracted and flat-fielded data are rectified using an arc lamp (for the
spectral direction) and an artificial point source (for the spatial
direction). To account for some spectral flexure between frames, we rectify
the data before sky subtraction. We correct for variations of the sky
background in the object and the sky frame by rescaling the count rate of the
sky frame to that of the corresponding object frame, carefully
masking the object, before reconstructing, aligning, and combining the
three-dimensional data cubes. Flux scales are obtained from standard star
observations taken once per hour at an air mass and position which match those
of the source. We also used stars to measure the spatial resolution. The
spectral resolution was measured from night sky lines. For a more detailed
description see, e.g., \citet{nesvadba06}. The data presented hereafter are those obtained with this careful reduction, but we checked that all features discussed in the paper are also seen in the dataset reduced with the standard SINFONI pipeline procedure.

\begin{figure}
\centering
\resizebox{8.5cm}{!}{\includegraphics{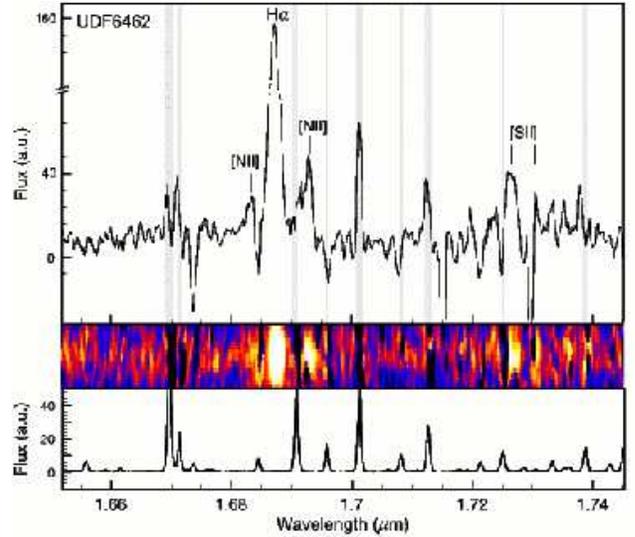}}
\caption{Integrated spectrum of UDF~6462 central regions (top). Sky lines are shown below. Shaded bands corresponds to wavelength ranges where the sky emission level exceeeds the average 1-$\sigma$ noise level estimated on the continuum. The HII line from UDF~6462 falls in a region free of telluric OH lines for at least $\pm 600$~km~s$^{-1}$ around the central wavelength. The [NII]$_{\lambda 6582}$ line is also unaffected by sky lines. The 2-D spectrum shown on the  middle corresponds to a pseudo-slit ranging from $\delta=-27^o$~47$'$~58.5$''$ (top) to  $\delta=-27^o$~47$'$~59.5$''$ (bottom); the pixel size on this plot is the raw SINFONI pixel size in both the spatial and spectral directions.}\label{f4}
\end{figure}

\subsection{Spectroscopic redshift and neighbouring systems}
We identified the H$\alpha$, [NII] and [SII] emission lines associated with UDF~6462 in the SINFONI datacube; the associated spectroscopic redshift is $z=1.571$. In UDF~6911, we detect the [OIII] and H$\beta$ emission; the associated redshift is $z=2.071$. As for UDF~6886, a single emission line is detected and assumed to be the H$\alpha$ line; in this case the redshift is $z=1.427$. Results on the internal kinematics of UDF~6911 are shown in Section~4.3. The emission line of UDF~6886 is not spatially resolved in our data. 

The integrated spectrum of UDF~6462 is shown in Fig.~\ref{f4}. The HII emission line from UDF~6462 falls far enough from telluric OH lines so that robust moment maps can be extracted. The detection of [NII] and [SII] emission is fainter and limited to the central regions. Because these lines are faint and/or contaminated by OH lines, associated velocity fields could not be properly retrieved over a significantly extended area.

The H-band continuum map of the system is shown in the third panel of Fig.~\ref{f2}. This continuum map is based only of the parts of the spectrum that are not affected by an OH sky line or another emission line. Linear interpolations were used to recover the continuum over the regions affected by these lines, and the continuum intensity was finally computed for wavelengths within the H-band range.

The two companions of the main system have largely different redshifts. In spite of their apparent proximity, similar colors and comparable clump sizes, these are not nearby systems in the 3-D Universe. The spurious apparent proximity and similarity of these systems illustrates that the identification of interacting/merging galaxies at high redshift without accurate spectroscopic redshifts cannot be rigorous. In the present case, the disturbed morphology of UDF~6462 could have further suggested an on-going interaction with its apparent companions, which spectroscopic redshifts rule out, showing that morphology is not always a good tracer of on-going galaxy interactions.

UDF~6462 is {\it not} interacting with its apparent neighbors; nevertheless this does not rule out a merger origin for this system itself. The blobs gathered into this irregular clump cluster may still be several independent units merging together instead of internal fragment of a single system, and this is what is studied in the following.

\subsection{Results: kinematics of the UDF~6462 clump cluster}

\begin{figure*}
\centering
\sidecaption
\resizebox{12cm}{!}{\includegraphics{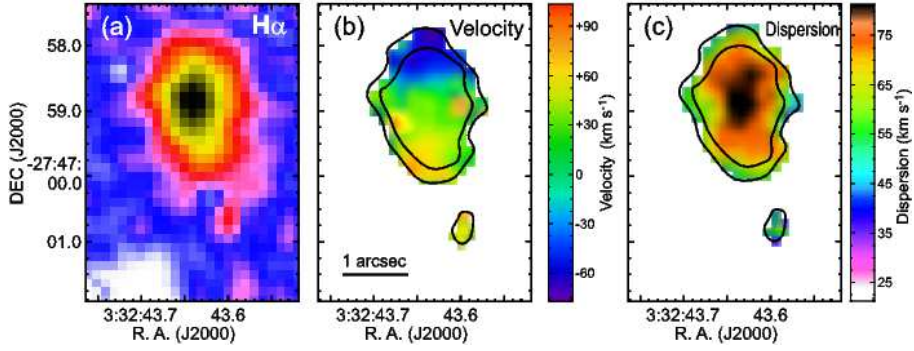}}
\caption{Moment maps for the H$\alpha$ line from our VLT/SINFONI data. {\bf (a)} $m=0$ intensity map, {\bf (b)} $m=1$ velocity field, and {\bf (c)} $m=2$ velocity dispersion field for UDF~6462. The  2-$\sigma$ and 5-$\sigma$ contours of the zero moment map are shown; the southern blob is detected at about 2.6-$\sigma$, with a continuum signal at the same location (Fig.~\ref{f2}). Instrumental broadening was subtracted from the dispersion map. UDF~6462 shows signs of rotation with a velocity gradient and a central peak in the velocity dispersion; however kinematical disturbances are larger than in classical low-redshift disk galaxies.
}\label{f5}
\end{figure*}

\begin{figure*}
\centering
\resizebox{5.5cm}{!}{\includegraphics{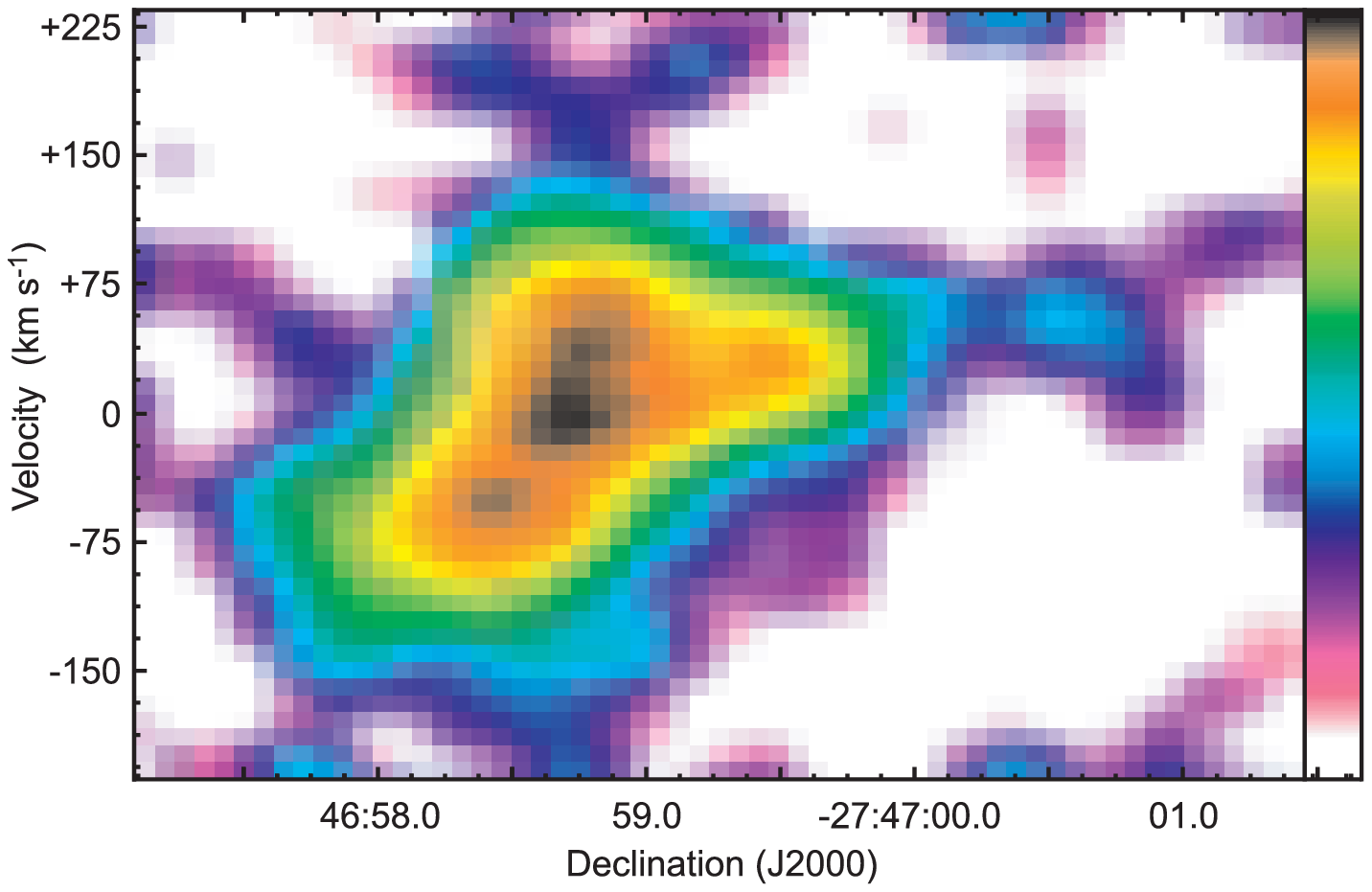}}
\resizebox{11cm}{!}{\includegraphics{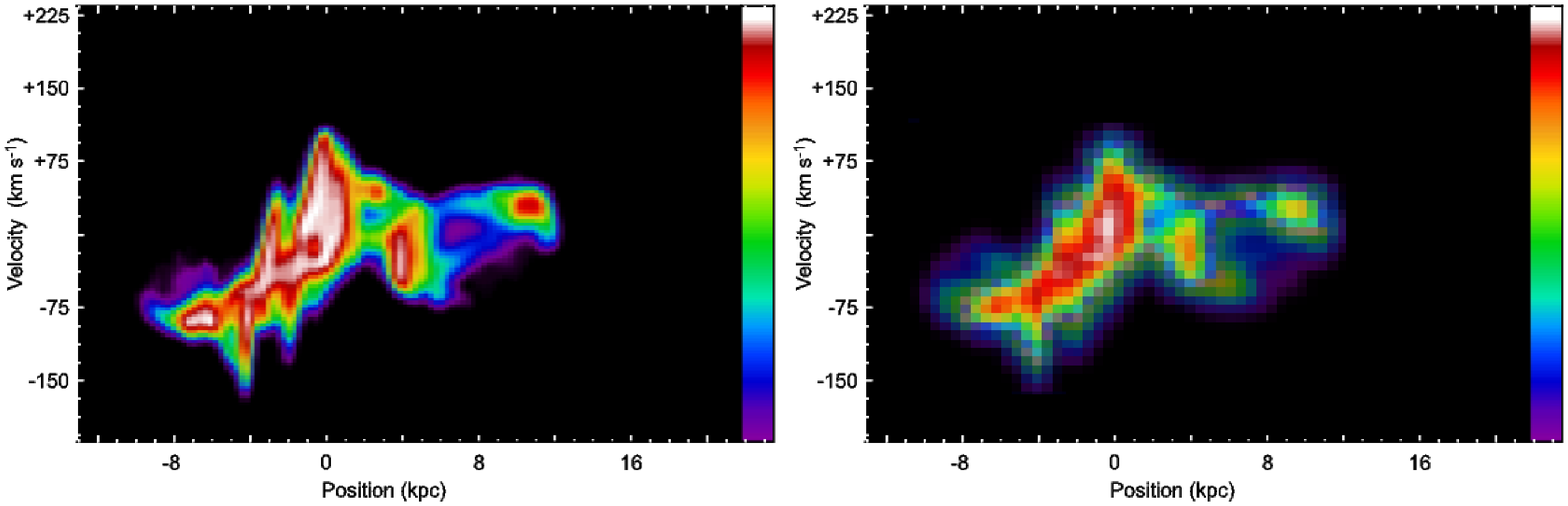}}
\caption{{\bf Left:} Position-Velocity diagram of UDF~6462 along its apparent major axis. The thick pseudo-slit used to compute this diagram is shown in Fig.~\ref{f2}, data have been resampled to a twice smaller pixel size on this P-V diagram. Note the global velocity gradient corresponding to the velocity field in Fig.~\ref{f5}, the steepest parts of the gradient around the emission peak corresponding to the peak in the line-of-sight velocity dispersions.
{\bf Right:} Position-Velocity diagram for the model in Section~3, along its projected major axis, at full resolution and degraded to the spatial seeing, pixel size and instrumental velocity resolution. A global velocity gradient as in UDF~6462 is reproduced, together with large local disturbances and rather high velocity dispersions. Each clump in this modeled clump-cluster has an internal velocity gradient associated with its own spin, which could not be resolved at the SINFONI resolution. 
}\label{f6}\label{sim:pv}
\end{figure*}

The first moment map associated with the H$\alpha$ emission line in UDF~6462 is shown together with the H$\alpha$ intensity contours in Fig.~\ref{f5}. The main regions are detected with a signal-to-noise\footnote{we measured the S/N ratio as the ratio of the flux in the HII line to the noise measured on the continuum, following for instance \citet{flores06} with a single line in our case.} (S/N) ratio above 3. The detection of the Southern blob is more uncertain, at about S/N=2.6, but this blob is also seen on the Position-Velocity (P-V) diagram shown in Fig.~\ref{f6}, on the continuum map (Fig.~\ref{f2}). The velocity field in the main region shows a large-scale velocity gradient with the Northern parts approaching and the Southern parts receding. The (suspected) Southern blob follows the same trend. 
One can however note that this velocity gradient is severely disturbed compared to what would be expected for a usual rotating disk. The $V/\sigma$ ratio\footnote{The $V/\sigma$ ratio here is not corrected for inclination because the inclination is unknown (or model-dependent), so it is not a direct indicator of the rotational support in this system.} is typically between 1 and 2. In particular, Fig.~\ref{f5} shows several clumps with velocities larger or smaller than expected for a purely rotating system. The P-V diagram along the apparent major axis of UDF~6462, shown in Fig.~\ref{f6}, confirms the overall trend for a velocity gradient throughout the system but with local disturbances; in particular a clump at $\delta$=-27$^o$47'59.4'' in the observation on Fig~\ref{f6} has a velocity lower than expected from the large-scale gradient, and a similar perturbation is found in the model at X=4~kpc in the same figure.

The kinematics of UDF~6462 thus shows some sign of rotation, with on average a velocity gradient along the major axis of the system. The velocity dispersion map shown in Fig.~\ref{f6} has a maximum in the dispersion around $\delta \simeq -27^o 46' 58.8''$ which is the region where the velocity gradient on the P-V diagram is maximal. This is a signature expected for rotation \citep[e.g.,][]{FS06, genzel06, puech06}, in particular because the high-dispersion region is extended in the direction perpendicular to the velocity gradient (i.e., along the morphological minor axis). Moreover, the highest slope in the P-V diagram lies on the continuum peak, which further indicates that this velocity gradient traces a rotation curve rather than some other kind of motion.

This underlying large-scale rotation pattern is, however, largely disturbed at the local scale of individual clumps. Some have velocities smaller or larger than expected for pure rotation, with discrepancies up to 40-50~km~s$^{-1}$. This is larger than the velocity anomalies caused by spiral arms and other internal structures in classical low-redshift disk galaxies \citep[e.g.,][]{rots75}. As a result, the velocity dispersions and local kinematical disturbances are high, nearly comparable to the rotation velocity if the large-scale gradient is tracing rotation.

\subsection{Metallicity and central enrichment in UDF~6462}

\begin{figure*}
\centering
\sidecaption
\resizebox{12cm}{!}{\includegraphics{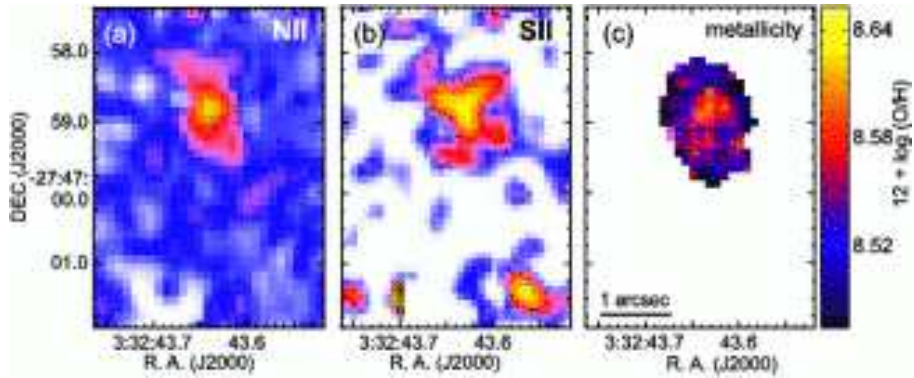}}
\caption{{\bf (a,b)} [NII] and [SII] intensity maps of UDF~6462. Metals appear to be centrally concentrated around the reddish proto-bulge and UDF~6462 shows sign of central enrichment illustrated by the metallicity map {\bf (c)} where the O/H ratio is estimated from the [NII]/HII flux ratio (see text).}\label{f10}
\end{figure*}

Intensity maps of the [NII] and [SII] emission lines are displayed in Fig.~\ref{f10}. The [NII] and [SII] emissions seem more centrally concentrated than the H$\alpha$ emission. Metallicity estimates, based on [NII]/H$\alpha$, within an aperture centered on the surface brightness peak in the H-band compared to that of the entire system, suggest that this region of high metallicity is real and not due to lower signal to noise in the outer regions of emission.

We used the N2 ratio, defined by $\mathrm{N2} = \log \left( \mathrm{NII}_{\lambda6584}/\mathrm{H}\alpha \right)$, as a proxy for the Oxygen abundance, as proposed by \citet{storchi-bergmann94}. \citet{pettini-pagel04} determined the relation:
\begin{equation}
12 + \log (\mathrm{O}/\mathrm{H}) = 8.90 + 0.57 \times \mathrm{N2}
\end{equation}

The N2 ratio in UDF~6462 supports star formation as the powering source of the optical lines \citep{erb06,halliday08}, consistent with the fact that this system does not show a mid-infrared excess typical of AGNs. The N2 ratio is also not as high as in could be for systems with strong winds \citep{LH96}. To derive the bulge metallicity, we measured intensities inside a 0.6\arcsec diameter aperture centered on the NICMOS H-band emission peak. The global metallicity was derived inside the 3-$\sigma$ contour of the H$\alpha$ flux map. We find:
\begin{itemize}
\item $12+ \log (\mathrm{O}/\mathrm{H}) = 8.59 \pm 0.04$ for the bulge 
\item $12+ \log (\mathrm{O}/\mathrm{H}) = 8.53 \pm 0.02$ for the whole system   
\end{itemize}
where the indicated uncertainties correspond to the estimated noise level in emission maps\footnote{Because our data are not flux-calibrated, these error bars are only related to the spatial variations of O/H; a constant additional offset may affect both measurements in the same way.}. This corresponds to a metallicity of $Z \simeq 0.5 Z_{\sun}$ for the whole system. The calibration of the N2 ratio might suffer from systematic uncertainties, but there is rather robust evidence for central enrichment in this clump-cluster system. The relative Oxygen abundance in the central redder blob is about 20\% higher than in the rest of the system. The whole system agrees with the mass-metallicity relation of $z\sim2$ established by \citet{erb06} and \citet{halliday08} and has a slightly lower metallicity than Luminous Infrared Galaxies at $z \simeq 0.6$ \citep{liang04}, but the bulge metallicity alone is significantly enriched and would be in between the $z\sim2$ relation and that of low redshift galaxies \citep[SDSS,][]{tremonti04}. The metallicity map obtained from the N2 ratio displayed in Fig.~\ref{f10} shows this central enrichement, and a radial metallicity gradient spanning from the central red bulge towards the rest of the system.


\section{Numerical Modelling}

Large kinematical disturbances are frequently interpreted as evidences for interaction/mergers, in particular when they are associated with complex morphologies (see Introduction). In the following, we use an $N$-body numerical model to study whether an internal origin for the star-forming clumps in UDF~6462 could be another viable explanation of such properties.

\subsection{Simulation technique}
The numerical technique used here is similar to that presented in more detail in BEE07 and references therein; we here only recall the main aspects. The N-body code models stellar, gaseous, and dark matter particles. The gravitational potential is computed on a $256^3$ grid using an algorithm based on Fast Fourier Transforms and optimized for vector computers. The grid cell size and gravitational softening length are 180~pc. We use $2\times 10^5$ stellar particles, $2\times 10^5$ gas particles, and $3\times 10^5$ dark matter particles. Star formation is accounted for using a local Schmidt law with an exponent 1.5 \citep{kennicutt98}. This 1.5 exponent is applied to the gas volume density in our models, which is equivalent to applying it to the surface density assuming a constant thickness, as is the case in our initial conditions. The dissipative nature of the turbulent interstellar medium is modeled with a sticky-particle scheme.

\subsection{Model parameters}
The physical parameters of the model presented here are close to that of model~6 in BEE07, except that a moderate initial asymmetry is introduced in the system. The baryonic disk initially has a flat radial profile (uniform surface density). This is because clump-cluster galaxies generally do not have a spiral-like exponential profile, but generally show irregular and less concentrated profiles \citep[e.g.,][]{EEVFF}, which is the case for UDF~6462 at least in optical bands. The hypothesis of a uniform surface density was made because it is a simple and not restrictive choice already made in BEE07 models, but is not required to produce a clump-cluster morphology -- clump interactions and migration will anyway rapidly redistribute the disk masss.

This disk has a radius of 8~kpc radius and a mass of $6\times 10^{10}$~M$_{\sun}$, with a constant scale-height of 700~pc. The initial gas fraction is 50\%, giving a stellar mass of $3\times 10^{10}$~M$_{\sun}$ \footnote{This is compatible with the observational estimate (see Sect.~2.1) once some more stars have formed in the simulation}. Stars are initially given random motions with a Toomre parameter $Q_\mathrm{s}=1.3$, and gas particles have a velocity dispersion of 11~km~s$^{-1}$. 

As the central blob of UDF~6462 is redder than the rest of the system and could be a primordial bulge, a spherical bulge of 14\% of the disk mass is added to the initial conditions in the model. This bulge fraction is compatible with observational estimates for UDF~6462 (15-20\% from the SED fits in Appendix~A), because 14$\%$ is the initial fraction in the model, which grows with time during the clump cluster evolution, and because the observed value is an upper limit to the actual bulge mass. Note furthermore that the bulge mass is not a crucial parameter to study the large-scale kinematics of clump cluster formed from fragmented disks: even bulgeless models in BEE07 show the same kinematical properties (disturbed rotation) as we find below for the present model.

The dark matter halo is modeled as a Plummer sphere of total mass\footnote{This total mass is that of the un-truncated sphere.} $2.9\times 10^{11}$~M$_{\sun}$ and scale-length 12~kpc, truncated at 24~kpc. The dark-to-visible mass ratio within the initial disk radius is 0.35. Models of unstable disk evolution with NFW dark matter profiles are presented in \citet{BHpaper}, and the different halo profile does not radically change the evolution of clump clusters formed this way.

\begin{figure*}
\centering
\resizebox{17cm}{!}{\includegraphics{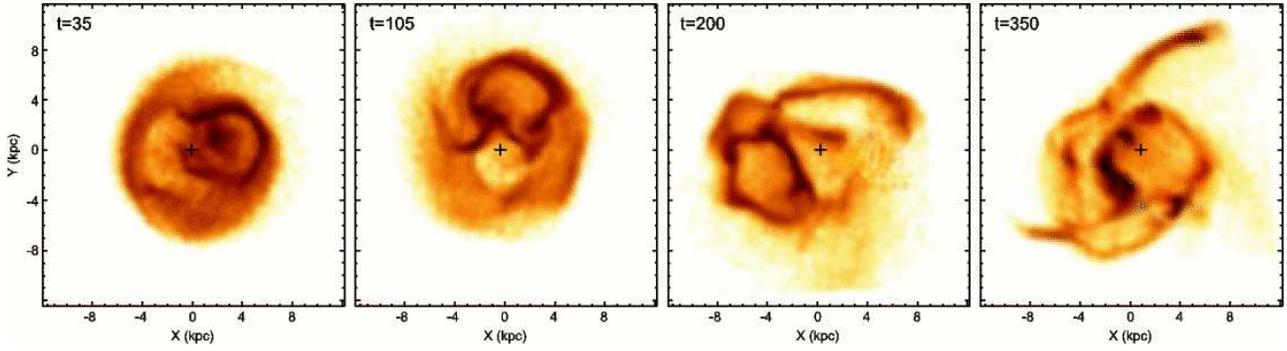}}
\caption{Snapshots of the disk mass density in the model, with time indicated in Myr. The halo mass center is marked by the black + symbol. The initial conditions are closely similar to the first snapshot, with a uniform disk density. The last snapshot shown here is slightly before the projection reproducing the UDF~6462 morphology (Fig.~\ref{sim:proj}). Note that the asymmetrical aspect resulting from the initial disk-halo offset is amplified by a long spiral arm extension and the outer location of one of the clumps.}\label{sim:snap}
\end{figure*}

Because the observed system is asymmetrical, the disk center is spatially offset from the halo and bulge mass centers by 10\% of this radius. Given that the degree of morphological lopsidedness in 
present-day spirals is frequently as large as 10 or even 20\% \citep[e.g.,][]{bournaud05m1, angiras07, reichard07} this is a realistic, moderate initial asymmetry\footnote{There are furthermore reasons to assume that $z \sim 1$ disks are more lopsided than $z \sim 0$. Be these asymmetries caused by interactions, by asymmetrical gas infall, or by halo asymmetries, these causes themselves should be more frequent/efficient at high redshift.}. The first snapshot in Fig.~\ref{sim:snap} illustrates the initial asymmetry.

\begin{figure*}
\centering
\sidecaption
\resizebox{12cm}{!}{\includegraphics{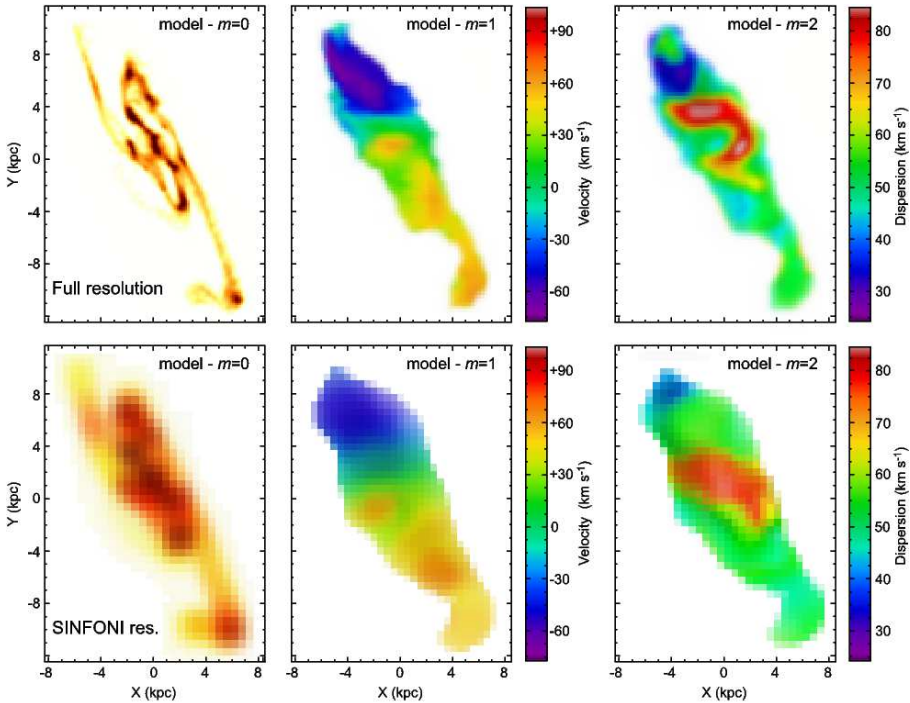}}
\caption{Projection of the model reproducing the UDF~6462 clump-cluster morphology, after $380$~Myrs of evolution. {\bf (a)} projected density map of stars formed less than 300~Myrs ago to mimic the V-band map shown in Fig.~\ref{f2}. {\bf (b)} line-of-sight velocity map. The velocity field was smoothed to a 1~kpc resolution (Gaussian smoothing) to reduce the numerical noise, and a cut-off along an isodensity contour was applied. {\bf (c)} line-of-sight velocity dispersion map. Under this projection, model qualitatively reproduces the severely disturbed rotation pattern observed in UDF~6462, with a large-scale velocity gradient and a central peak in the velocity dispersion. {\bf (d,e,f)} Same moment maps degraded to the spatial resolution of our SINFONI observations. Irregularities in the velocity field are still visible, for instance the high-velocity region around X=-2 and Y=-1 kpc, which is comparable to the velocity anomalies observed in UDF~6462.
}\label{sim:proj}
\end{figure*}

\subsection{Morphology}

The gas-rich, unstable disk in the present model is similar to the one studied in BEE07 and follows a similar evolution. The disk fragments into massive clumps in about $10^8$ years, the formation of which can be followed in Fig.~\ref{sim:snap} and \ref{sim:proj}. This system forms eight big clumps embedded into a fainter underlying disk; the mass fraction in the clumps reaches a peak value after $t=270$~Myr when 41\% of the disk mass is gathered in the clumps. 

Because of the initial asymmetry, the clump distribution itself can appear quite asymmetric. Depending on the projections, the clumpiness can increase the apparent asymmetry. For instance, the last instant in Fig.~\ref{sim:snap} would have a high apparent asymmetry seen from the bottom-right, and a lower asymmetry seen from the bottom-left. We also note the development of an lopsided spiral structure from the initial asymmetry and instability. This structure resembles the Southern extension of UDF~6462, which might thus be formed without an external tidal interaction.

The instability of this massive gas-rich disk, combined with its moderate initial asymtery, results in the quite asymmetrical and clumpy projection shown in Fig.~\ref{sim:proj} with an inclination of 64 degrees. This projection is morphologically similar to the HST V-band image of UDF~6462 shown in Fig.~\ref{f2}.

\subsection{Kinematics}

We show in Fig.~\ref{sim:proj} the velocity field of the modeled clump cluster for the projection which looks similar to UDF~6462. The velocity field shows a large-scale velocity gradient, but local variations and irregularities can be seen which resemble those observed in the SINFONI velocity field of UDF~6462. In addition, the P-V diagram along the apparent major axis is shown in Fig.~\ref{sim:pv}, both at full resolution and at the SINFONI resolution. Clumps on this P-V diagram show an internal spin that could not be detected at the SINFONI resolution. On larger scales, clumps follow a global velocity gradient associated with the rotation of the system, but the individual velocity of each clump can be offset from this global gradient by a few tens of km~s$^{-1}$. Similar kinematical irregularities are also shown for some cases in BEE07, and are caused by clumps interacting together and migrating in the system. In the present case, the P-V diagram of the system also incorporates the asymmetry, which gives it an overall aspect qualitatively similar to the observed one shown in Fig.~\ref{f6}. A velocity gradient can be detected and traces the overall rotation of the system, but the internal evolution of the clump cluster causes significant disturbances to the kinematics, that are larger than the disturbances caused by spiral arms and internal structure in regular spiral disks \citep[e.g.,][]{rots75}.

The observed peak-to-peak velocity extension of the P-V diagram of UDF~6462 is about 120~km~s$^{-1}$. Assuming that the circular velocity is traced by the envelope at 50\% of the peak level \citep[e.g.,][]{donley06}, the projected circular velocity $V_{\mathrm{c}} \sin i \simeq 90 $~km~s$^{-1}$. Assuming an inclination $i=63 \deg$, as suggested by our numerical model and consistent with the observed morphology, this corresponds to a circular velocity of about 100~km~s$^{-1}$.
\bigskip

Both the morphological and kinematical properties of UDF~6462 are qualitatively reproduced by the clump-cluster model presented here, at the instant and under the projection that we have chosen. We do not claim that the real system has followed exactly the same evolution in particular, because we simply picked a similar-looking model among several ones, and the solution is probably not unique. Nevertheless, the present model illustrates the fact that the clumpy morphology and the disturbed kinematics of UDF~6462 can be reproduced by models based on the internal evolution of unstable gas-rich disks: this model shows that the morphology, clumpy and asymmetric, and the kinematics with local perturbations, can both be accounted for by the internal evolution of gas-rich disks. To achieve this, we have here added a moderate initial asymmetry in the present model, which under a chosen projection reproduces the peculiar morphology observed for this system and its disturbed morphology at the same time. Other models presented in BEE07 show similar morphologies and kinematical disturbances of the same order, even if not reproducing the individual case of UDF~6462.


\section{Discussion}
\subsection{UDF~6462: Unstable disk or multiple merger?}

The UDF~6462 clump cluster has been considered a good candidate for a merging system, because of its irregular and asymmetrical morphology compared to some other chain galaxies that have more symmetrical aspects. In such a group model, initially separate dark matter haloes with their stellar and gaseous content each would be merging together, and be presently observed in a phase resembling a compact group. It could also be a binary merger with several star forming clumps.

However, our numerical model shows that such a complex morphology can result from the internal evolution of an unstable gas-rich disk galaxy. In this fragmentation model, a single dark matter halo contains or accretes large amounts of gas, so that the primordial disk can become Jeans-unstable and fragments into several large star-forming clumps, taking the appearance of a clump-cluster. Only a moderate (realistic) initial asymmetry is needed in the model to qualitatively reproduce the projected morphology of UDF~6462. The kinematic data obtained with SINFONI unveil a large-scale velocity gradient, together with local disturbances. Such a large-scale velocity gradient is a signature expected for a clump cluster formed by internal instabilities in a disk galaxy. The irregularities in the velocity field are then explained by clump-clump interactions that cause the individual velocity of each clump to differ significantly from the initial rotation velocity. Both the morphology and the kinematics of UDF~6462 can thus be accounted for by an unstable disk scenario.

The viability of this first scenario does not strictly rule out a merger origin. The blobs observed in UFD~6462 could be initially independent sub-units merging together at the same time to form a more massive galaxy. However, several observed properties appear unlikely under this hypothesis:
\begin{itemize}
\item (1) First, the clumps have comparable masses and sizes (see Figs.~\ref{f2}~and~\ref{f11}) which is typical of most chains and clump cluster galaxies. The dynamical friction timescale for merging galaxies is inversely proportional to the mass. Blobs that have the same mass as the most massive one should merge rapidly, and low-mass blobs should survive longer. The present morphology of UDF~6462 could then only be a transient one, unlikely to be observed. A massive blob surrounded by several lower-mass ones would be a more likely configuration. On the other hand, the similar masses for all clumps are naturally expected for the internal fragmentation of a primordial disk, because the typical clump mass in this case would be the Jeans mass in the outer disk. 

\item (2) The large-scale rotation-like velocity gradient and the central peak in the velocity dispersion are also characteristics of rotating disks that are predicted to be preserved (with disturbances) in a clump cluster formed from an unstable disk. Such properties are not necessarily expected for a merging system. In particular, there is no reason for which merging blobs would be globally rotating in the same direction. The observed velocity dispersions are rather high but do not dominate the large-scale velocity gradient; for a multiple merging they would more likely be larger the rotation velocity. Actually the merging timescale is faster for systems that rotate in the same direction, because of angular momentum removal in tidal tails, while counter-rotating systems take longer to merge. The absence of counter-rotating unit among the several blobs is a possible but unlikely property for a multiple merger origin.

\item (3) The radial metallicity gradient (Fig.~\ref{f10}) with central enrichment in the reddish bulge-like blob is also expected for a (clumpy) disk, and the total metallicity is about that expected for a disk of this mass. The centrally concentrated [NII]/H$\alpha$ ratio is similar to what \citet{FS06} found in a rotating disk at $z\simeq 2$ (Q2343-BX610), with a ratio between the central metallicity peak and the outer disk metallicity larger by 35\% but comparable to what we find here in UDF~6462. Such a disk-like metallicity distribution would be another coincidence if UDF~6462 is an-going group merger. Generic models of galaxy mergers and disk fragmentation in Appendix~B confirm that a radial metallicity gradient is better preserved by the disk fragmentation mechanism than in early-stage galaxy mergers with clumpy morphologies.

\item (4) Stellar population studies confirm that the clumps are young, star-forming objects around an older, central, bulge-like object. The observed SEDs and theoretical fits for the whole galaxy, the bulge, and several clumps are shown in the Appendix~A and Fig.~\ref{f11}. The central region that we have called bulge is clearly redder than the clumps. The reddest clump is an extended region jutting off from the bulge and may be part of the bulge. The central reddish bulge has an age around 1~Gyr. The clumps have an age of $\sim50$ Myr with a comparable decay time in the star formation rate. There appears to be no significant old stellar component in the clumps other than the central bulge: if these clumps were separate galaxies in a merging process, they should have older stellar components in them, formed before this merger as in any galaxy. The absence of old stars in the clumps further supports our model in which they formed recently by the fragmentation of a gas-rich disk. Moreover, these clumps appear to have similar ages (similar SED slopes) which again arises naturally in the disk fragmentation scenario.


\end{itemize}

The observed properties of UDF~6462 are thus consistent with an unstable-disk origin, while they would be unlikely coincidences in a merger origin. We also note that a stellar mass of $3.3\times 10^{10}$~M$_{\sun}$ and an external diameter of $\sim 10$~kpc (without the low-mass Southern extension) would place the system in agreement with the mass-size relation of disk galaxies from \citet{trujillo06} which, at a slightly higher redshift of 1.7, indicates an {\it half-light} diameter of 6~kpc for a galaxy of this mass. Furthermore, the 5~kpc radius and 100~km~s$^{-1}$ circular velocity inferred above are consistent with the size-velocity relation observed for star-forming disks at z=1.5-2 by \citet{bouche07}, while submillimeter galaxies (likely major mergers) are found to have much lower size:velocity ratios by these authors.

These various properties cannot definitely rule out a merger origin, also because only one system with direct evidence for massive clumps from HST/ACS and NICMOS imaging has been studied so far. However, for the merging of a group of initially separate galaxies, one would expect a concentration of dark matter around each visible blob, so that the largest velocity gradients should be internal to the clumps, and not in the form of a global rotation of the system. One might assume that, in a merger model, the initially separate dark haloes could have merged into a single one, while their baryonic counterparts would still be separate. However such a system should resemble Local Compact Groups, in which the prominent kinematical signatures are internal gradients inside each galaxy with little global rotation of the group \citep[e.g.,][and references therein]{amram07}. Local systems have lower gas fractions, but tidal interactions in the early stages of mergers are dominated by the DM and stellar potential, so that the kinematic response of the gas would at first order be similar in higher redshift compact groups. This comparison further confirms that the properties of UDF~6462 are unlikely for a merger mechanism. On the other hand, the fragmentation model within a single dark matter halo predicts (see the simulation here and those in BEE07) that clumps have some internal rotation, but with low velocities (a few tens of km~s$^{-1}$), which could not detected at the SINFONI resolution (see Fig.~\ref{sim:pv}). The velocity dispersions are rather high for a disk (compared to regular spirals), with $V/\sigma \simeq $1.5-2 in the observations and $V/\sigma \simeq $ 2.5 in the model. The small difference is likely explained by the larger beam size in observations. Models of galaxy mergers usually result in much lower $V / \sigma$ ratios, smaller than 0.5 in the case of multiple mergers \citep[e.g.,][]{bournaud07}.

UDF~6462 is presently a quiescent galaxy. Its star formation rate to stellar mass ratio $SFR/M_\star \simeq 1.5$~Gyr$^{-1}$, which is consistent with the typical $SFR-M_\star$ relation found for galaxies of this mass by \citet{daddi07} at $z \sim 2$. Stellar population studies in EE05 however suggest that the star formation activity was somewhat stronger in the last few $10^8$~yr. The formation of massive clumps in gas-rich disks can indeed trigger star formation: models in BEE07 typically reach star formation rates of 30--40~M$_{\sun}$~yr$^{-1}$, up to ten times higher than the progenitor axisymmetric disks. Immeli (2004a) reports even higher star formation rates of 200~M$_{\sun}$~yr$^{-1}$ but in the more extreme case of initially gas pure disks nearly as massive as the Milky Way. The lifetime of these systems in BEE07 is about 0.5--1~Gyr, which is actually a lower limit because regulation by supernovae feedback was not included in these models. The circular velocity and dispersion estimated above corresponds to a dynamical (total) mass of $\sim 4-5 \times 10^{10}$~M$_{\sun}$ within the optical extent of the system, consistent with the stellar mass (Section~2.1) and implying a rather moderate ($\sim$20\%) gas fraction at the observed instant\footnote{assuming that the mass within the optial radius is mostly visible with little contribution from the halo, as is generally the case for Local disk galaxies}. However, an SFR of $\sim 50$~M$_{\sun}$~yr$^{-1}$ and an age of $\sim 3-4 \times 10^{8}$~yr imply an initial gas fraction of about 50\%, so that these mass estimates are compatible with the starting conditions of our model. We also note that this 50\% initial gas fraction in the model is in agreement with observations by \citet{erb06b} and \citet{daddi08} and assumptions in \citet{bouche07}. Furthermore, models with different initial gas fraction in BEE07 also show disturbed rotation comparable to the present case.

Although the present models are initially asymmetric and highly complex structures develop quickly, real galaxies could be even more complex, especially in the off-plane dimension, as a result of active inflow from cosmic streams and previous mergers.
However, the model clearly indicates how internal clumpiness in a gas-rich galaxy can affect the kinematics. The comparison with observed properties then shows that the scenario in which the UDF~6462 clump cluster would have formed by internal instabilities in a gas-rich disk is a fully plausible one, in spite of the complex morphology and disturbed kinematics. These observed properties can be explained, and the global rotating expected in this scenario is observed as a large-scale velocity gradient throughout the clump cluster.

\subsection{Kinematics of the bent-chain UDF~6911}

\begin{figure}
\centering
\resizebox{6cm}{!}{\includegraphics{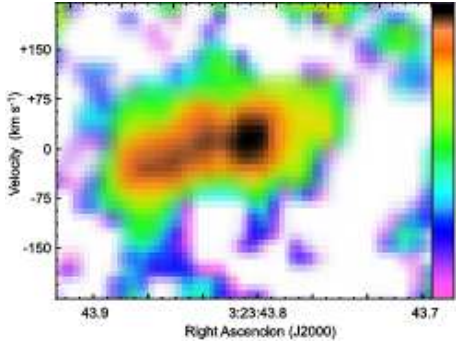}}
\caption{Position-Velocity diagram of UDF~6911 along its apparent major axis, in the [OIII] emission line. The thick pseudo-slit used to compute this diagram is shown in Fig.~\ref{f2}.}\label{6911}
\end{figure}

The Eastern companion of UDF~6462, UDF~6911 (Fig.~\ref{f2}), has the elongated morphology of a bent chain galaxy at $z=2.07$. The same question on its origin can be posed as for UDF~6462. The kinematics of this system is only poorly resolved in our SINFONI data. The P-V diagram of the [OIII] emission line along the apparent major axis of this object is shown in Fig.~\ref{6911}. Here again, we detect a velocity gradient with a shape compatible with the flat rotation curve expected for a disk galaxy. Assuming that the circular velocity is traced by the envelope at 50\% of the peak level \citep[e.g.,][]{donley06} to account for beam-smearing effects, and that this system is observed close to edge-on, the rotation velocity is estimated to be about 75~km~s$^{-1}$. The resolution is too low to unveil local disturbances, but at least the global rotation pattern expected for a fragmented disk is found here, indicating that this interpretation is fully possible for this second object as well as for UDF~6462.

\subsection{Clump clusters and spiral disk and culge progenitors, and their role in the star formation history}

Within the plausible scenario where UDF~6462 and UDF~6911 (and possibly other clump clusters and chain galaxies) formed by internal fragmentation of clumpy disks, primordial galactic disks have to accrete enough mass at high redshift so that large fragments of their disks can collapse though gravitational instabilities. The presence of large gas reservoirs around disk galaxies at $z \sim 1.5$, unveiled recently by \citet{daddi08}, shows that large amounts of gas representing at least half of the total baryonic mass can indeed be accumulated around primordial galactic disks. Massive disks with such high gas fractions are then likely to be gravitationally unstable and fragment into clump-cluster galaxies, or chain galaxies when seen edge-on, clump-clusters being face-on counterparts of the later as already suggested by \citet{EEH04}.

Fragmentation of gas-rich disks into clump clusters can then be an efficient mechanism to trigger star formation, with an efficiency at least comparable to that of major merger, if not higher, and longer star-formation timescales. Indeed, numerical models of merger-induced starbursts indicate that, on average, these have only modest intensities and durations \citep{DM07} even if some particular cases can be very efficient \citep[e.g.,][]{cox06}. Recent observations further suggest that galaxy interactions and mergers have a modest influence on the cosmic star formation history \citep{jogee08} and that internal processes in gas-rich disks play a more important role \citep{daddi08}. In a sample of Luminous Infrared Galaxies at $0.4 < z < 1.2$, \citet{zheng04} have identified a majority of disk and irregular galaxies and a relatively small fraction of mergers.

Later on, the clumps migrate towards the central region, being also progressively disrupted by the tidal field and their mutual interactions and redistributing the underlying disk mass into an exponential disk, as shown by numerical simulations \citep[][BEE07]{carollo07}. Bulges with masses up to $\sim$30\% of the baryonic mass can be formed, or grown if already present, by the central merging of clump remnants; the bulges formed this way have the properties of the classical bulges of spiral galaxies \citep{BHpaper}. Associated central black holes can be accreted at the same time from intermediate mass black holes formed by stellar collisions in the core of clumps, possibly explaining the black-hole-bulge scaling relations without mergers \citep[see][for the case of mergers]{johansson08}. Clump cluster and chain galaxies, when formed by the fragmentation of gas-rich primordial disks, are thus possible progenitors of present-day bright disk galaxies, as already suggested by \citet{vdbergh96}. 

Clump clusters were found by BEE07 to form a thick disk of old stars, together which a thinner disk of gas and younger stars, which can further grow in mass by continued accretion. One can notice that the rotation velocity found in UDF~6462 is lower than what would be expected for the stellar mass from the Tully-Fisher relation \citep{conselice05}. As the thick disks of spiral galaxies are observed to have a lack of rotation compared to thin disks \citep{YD05, YD06}, clumpy evolution at high redshift is also a viable formation mechanism for thick disks, and indeed high redshift disks are observed to be thick (Reshetnikov et al. 2003, Elmegreen \& Elmegreen 2006b).


\section{Conclusion}

The UDF~6462 clump-cluster galaxy at $z=1.57$ has a complex asymmetrical morphology, dominated by massive star-forming clumps. Spectroscopic observations of the H$\alpha$, [NII] and [SII] emission lines of this system using SINFONI on UT4 at ESO/VLT have revealed a velocity gradient throughout the system, possibly tracing global rotation with a circular velocity of about 100~km~s$^{-1}$. A central peak in the velocity dispersion further suggests that this velocity gradient likely traces rotation. This is expected if this clump cluster formed by internal fragmentation of an unstable gas-rich primordial disk. The velocity field shows large irregularities and dispersions which would be an anomaly for classical spiral disks, but can be explained by interactions between clumps, as shown by our numerical model. The unstable-disk model presented here reproduces, under a chosen projection, both the main morphological and kinematical properties of UDF~6462. 

An internal origin for clumpy disks is thus fully compatible with observed morphological and kinematical properties in the case of UDF~6462. This does not definitely rule out a merger origin for this clump cluster. Nevertheless, the rotation-like velocity field would not be necessarily expected this way, while it is predicted for internal instabilities in a gas-rich disks (BEE07). Furthermore, UDF~6462 shows signs of central enrichment in a bulge-like central condensation, and a global metallicity in agreement with the estimated disk mass. This also is consistent with internal clumpiness in a massive, gas-rich disk, and would be an unlikely coincidence for a merging system. A similar large-scale rotation-like velocity gradient is also found, although poorly resolved, in the nearby bent chain galaxy UDF~6911 at redshift $z=2.07$.

Just like morphological asymmetries and clumpiness, large kinematical disturbances of several tens of km~s$^{-1}$ can be a natural consequence of the internal evolution of gas-rich primordial disks if they become unstable and fragment into large star-forming clumps once they have accreted enough mass. Spectroscopic data should be obtained on more clumpy galaxies, and a systematic comparison with model predictions should be performed, before general conclusions on the origin of chain and clump cluster galaxies can be obtained. At least, our present results show that even a severely disturbed case like UDF~6462 could actually be formed by internal evolution and that hierarchical mergers are not necessarily needed to explain the formation of such systems. Highly disturbed morphologies and/or disturbed kinematics with high velocity dispersions have often been attributed to mergers in the recent literature. We have here shown that such properties do not provide robust evidence for mergers, and may have an internal origin as well. {Resolved spectroscopic data are needed to really constrain the origin of clumpy galaxies at high redshift on large samples, as well as large sets of numerical models to check that systematic kinematical classifications as in \citet{flores06} or \citet{shapiro08} can distinguish fragmented disks from merging systems.

If clumpy galaxies are formed by large instabilities in gas-rich primordial disks, as is likely the case for UDF~6462 and possibly also for UDF~6911, they can show large disturbances in their velocity field from one clump to the next one, but models presented here and in BEE07 predict that on the largest scales, velocity gradient tracing the initial rotation should be observed together with high velocity dispersions. High turbulent speeds observed in high-redshift disks \citep{genzel06, FS06} are in agreement with this scenario, as they increase the Jeans mass and enable the formation of large, massive clumps in gas-rich disks. Moreover, UV and optical star forming disks at high-redshift have high angular momentum compared to local disks \citep{bouche07}, which phases of clumpy evolution can contribute to dissipate rapidly while fueling bulges and shaping more concentrated disks with exponential profiles. The spectroscopic data presented in this paper thus support the hypothesis by Noguchi (1999), Immeli et al. (2004a,b), Elmegreen et al. (2005) and BEE07 that gas-rich primordial disks internally evolve through a clumpy phase into bright early-type disk galaxies with a massive exponential disk, a classical bulge, and possibly a central black hole.


\appendix

\section{Stellar population study of UDF~6462 bulge and star-forming clumps}

\begin{figure}
\centering
\resizebox{8cm}{!}{\includegraphics{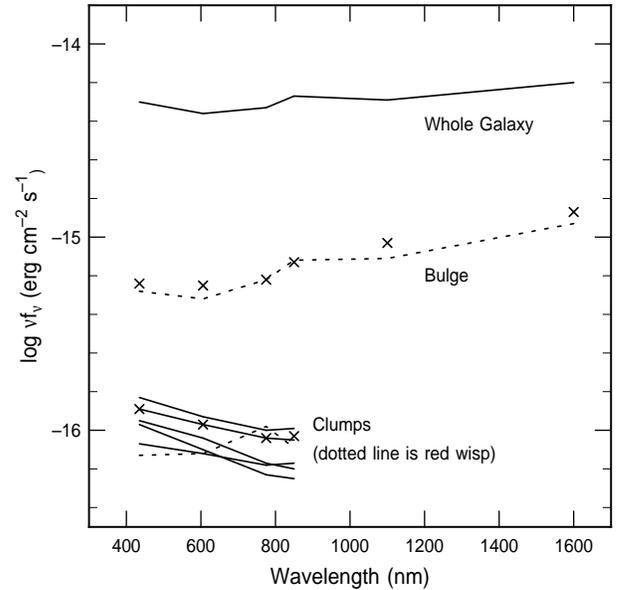}}
\caption{Spectral energy distributions of the whole system (top line), central
bulge-like object (middle line), and several clumps (bottom left). 
Symbols are the best model fits for the bulge
and for one of the clumps. A good fit is obtained for the clumps without
an old stellar population, while the bulge has a $\sim1$ Gyr old
population. The 'red wisp' is a small and somewhat redder clump at
($\alpha$=3$h$ 32$m$ 43.68$s$ -- $\delta$=-27$^o$ 46$'$ 58.8$''$) close to the bulge itself.
The typical uncertainties on the clump individual fluxes are 0.2 on $\log (\nu f_{\nu})$ in the optical and 0.3-0.4 in the NIR, so that the difference between the central bulge and the other clumps is robust.
}\label{f11}
\end{figure}

The observed SEDs for the whole UDF~6462 galaxy, the bulge, and several clumps are shown in figure \ref{f11}, using NICMOS resolution for the bulge and ACS resolution for the clumps. The bulge is clearly redder than the clumps. The reddest clump is an extended region jutting off from the bulge and may be part of the bulge. Theoretical fits to the SED of one of the clumps and to the bulge SED (crosses) are shown; details of these models are in EE05. Extinction curves from \citet{calzetti} and \citet{leitherer} were used, and intergalactic absorption by Lyman lines was included following \citet{madau}. The basic population evolution models are from \citet{bruzual} assuming a metallicity of 0.008, which is 0.4 solar, and a \citet{chabrier} IMF. Star formation is assumed to begin at some time prior to the redshift of the object and to decay exponentially with a characteristic time scale. The age of the region and the decay time, along with the extinction, are parameters of the fit. The best fits here require extinctions of $A_V=0.6$ mag for the clump and $A_V=1.2$ mag for the bulge. The models suggest that the central, reddish, bulge-like object has an age of around 1~Gyr (or even older if the actual extinction is the same as for the clumps), and the clumps have an age of $\sim50$ Myr with a comparable decay time in the star formation rate. There appears to be no significant old stellar component in the clumps. If these clumps were separate galaxies in a merging process, they would have older stellar components in them and look redder. No other part is as red as the central bulge or the reddish wisp near it (see also Fig.~\ref{f3}), which further suggests that no clump is a separate galaxy. The bulge is bluer than bulges in moderate redshift galaxies \citep[e.g.,][]{zheng04}, which suggests some on-going star formation; however the relatively poor resolution of NICMOS compared to the bulge clump observed with the ACS allows younger structure to contaminate the bulge SED, which would bias the bulge color slightly towards the blue.

\section{Metallicity distribution in galaxy mergers and fragmented disks }
The models from BEE07 used to interprete the kinematics of UDF~6462 do not directly provide predictions on the gas metallicity. To support the interpretation of the observed metallicity distribution in Fig.~\ref{f10} and Section~4.1, we show some Tree-SPH models including chemical evolution from \citet{DM07}, for clumpy star-forming systems formed by galaxy mergers and internal disk fragmentation.

\begin{figure}
\centering
\resizebox{8cm}{!}{\includegraphics{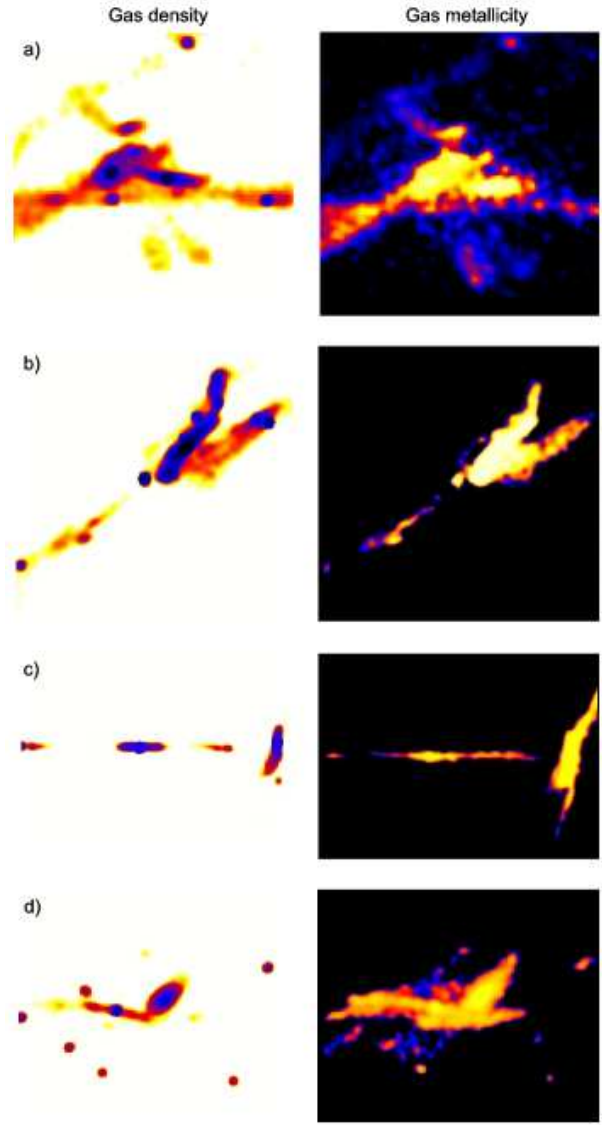}}
\caption{
Gas density and metallicity maps in several galaxy merger models with a clumpy gas distribution. Outer star-forming regions can have been previously enriched in the inner disks of the progenitor spirals, like the tail to the bottom-left in case b, the clump to the left in case c, or the various external clumps in cases a and d. These cases are typical of the merger models from the {\sc Galmer} database. Snapshots show between 25 and 40~kpc in radius and are all in log scale, with different intervals for each one as they represent different systems.}\label{metal_merg}
\end{figure}

\subsection{Galaxy merger models}

The galaxy merger models shown here are from \citet{DM07} and the {\sc Galmer}\footnote{available at {\tt http://galmer.obspm.fr}} database (Chilingarian et al. in preparation). The merging disk galaxies start with a realistic radial metallicity gradient in their disk. These binary mergers often have two main star forming regions at the center of each galaxy, but massive star forming regions also form in at larger radii by instabilities in spiral arms or in tidal tails \citep[see e.g.,][]{elm93}, sometimes giving a clump-cluster morphology in the gas distribution. The four cases shown in Fig.~\ref{metal_merg} are early stage mergers that have been selected in the database for their clumpy morphology without any criterium related to the metallicity. They are representative of the majority of the $\sim 900$ models available in the database. 

As shown in Fig.~\ref{metal_merg}, clump-cluster-like systems formed in early stages of galaxy mergers do not have a well organized metallicity gradient. In UDF~6462, the metals are observed to be centrally concentrated around the red proto-bulge and star formation takes place in the outer clumps. At the opposite, in these galaxy mergers models the star forming knots in the outer regions are associated to metallicity peaks. Even the outermost large star-forming regions in tidal tails around colliding galaxies, the TDG progenitors, are observationally known to have higher metallicities than classical disks outskirts, with quite often nearly-solar metallicitities \citep[e.g.,][]{duc2000}. Indeed, star-forming regions at large radii in mergers can have been previously enriched in inner regions and expelled by tidal forces. Hence, the radial metallicity gradient typical of disks is not preserved in the early phases of galaxy mergers (at least until the final relaxation) even when these have gas distribution similar to clumpy galaxies.

\subsection{Disk fragmentation models}

\begin{figure}
\centering
\resizebox{7cm}{!}{\includegraphics{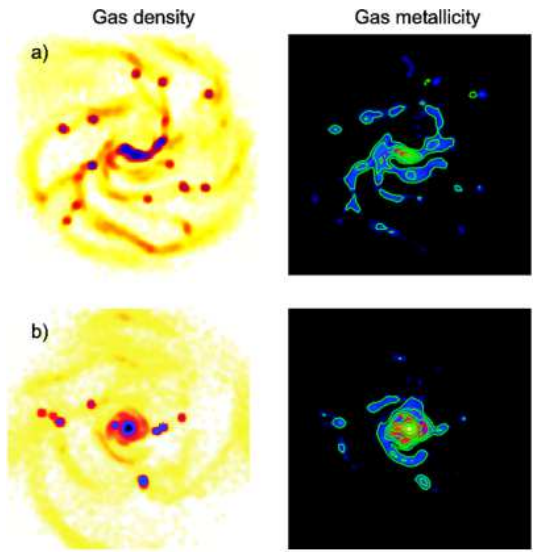}}
\caption{
Clump-cluster model formed by internal disk fragmentation. The gas density and metallicity maps are shown at two stages in the evolution. Green contours on the metal maps are metallicity contours, linearly spaced from 0 to 0.05. Each box is $40\times 40$~kpc. The star forming clumps have a moderate contribution to the metallicity distribution. The radial gradient from the progenitor disk is better preserved than in the merging of initially independent galaxies. 
}\label{metal_disk}
\end{figure}

The model shown on Fig.~\ref{metal_disk} is a gas-rich unstable disk, that gets fragmented in the same process as our dedicated model reproducing the kinematics of UDF~6462. Clumps form stars and contribute to the ISM enrichment, they interact and migrate inwards. Despite, a prominent metallicity peak remains clearly associated to the mass center, and no major metallicity peak is associated to the outer star forming clumps.
The radial metallicity gradient typical of classical disks, and observed in UDF~6462, is hence best preserved if clump cluster galaxies form by the fragmentation of gas rich disks than in the merger of independent galaxies.


\begin{acknowledgements}
We thank the ESO Director General for allocating DDT time to this observing program. The numerical calculations were carried out on the vector computer NEC-SX8R of CEA/CCRT and supported by the Horizon project. Useful discussions with Nicolas Bouch\'e, Pierre-Alain Duc, Fran\c{c}oise Combes and Chris Conselice are gratefully acknowledged. Suggestions by the referee and professional handling of the manuscript by the editor are appreciated.
\end{acknowledgements}






\end{document}